# Room-temperature ferroelectrically switchable quantum geometry in few-layer $WTe_2$ for complementary in-memory computing

Ruihan Wang[1,2†], Pengfei Wang[3†], Haoyun Chen[1], Yunze Peng[4], Bingyan Liu[1], Junlin Xiong[3], Xueyuan Zhang[5], Chen Pan[5], Xin Chen[1], Shengyuan A. Yang[6], Shi-Jun Liang[3]*, Feng Miao[3]*, Peng Song[1,2]*

[1]School of Electrical and Electronic Engineering, Nanyang Technological University, 50 Nanyang Ave, Singapore, 639798, Singapore

[2]CINTRA CNRS/NTU/THALESUMI 3288 Research Techno Plaza, Singapore 637553, Singapore

[3]Institute of Brain-Inspired Intelligence, National Laboratory of Solid State Microstructures, School of Physics, Collaborative Innovation Center of Advanced Microstructures, Jiangsu Physical Science Research Center, Nanjing University, 210093, Nanjing, China

[4]School of Physics, Beihang University, Beijing 100191, China

[5]Institute of Interdisciplinary Physical Sciences, School of Physics, Nanjing University of Science and Technology, Nanjing 210094, China

[6]Research Laboratory for Quantum Materials, Department of Applied Physics, The Hong Kong Polytechnic University, Kowloon, Hong Kong, China

*Corresponding authors: S. L. (sjliang@nju.edu.cn); F. M. (miao@nju.edu.cn); P. S.(peng.song@ntu.edu.sg)

†These authors contributed equally to this work.

**Abstract**

Quantum geometry, describing the inherent geometric structure of electron wavefunctions in momentum space, transcends the traditional charge degree of freedom and provides a novel physical basis for information encoding and processing. The key to such new computing paradigms is the non-volatile electrical programming of quantum geometric states at room temperature, which, however, has not been established. Here, we demonstrate ferroelectrically switchable quantum geometry in few-layer $WTe_2$, which uniquely enables complementary convolutional processing. By employing the intrinsic coupling between ferroelectric polarization and quantum geometry in few-layer $WTe_2$, we show that the second- and third-order nonlinear anomalous Hall effects (NLAHE) can be deterministically and electrically switched in a nonvolatile and correlated manner. The switching is robust at room temperature for ~$10^4$ cycles and retention of ~$10^5$ s. Furthermore, leveraging the opposite switching behaviors of second- and third-order NLAHE at room temperature, we demonstrate complementary in-memory computing and implement a hardware-level complementary convolution kernel. This kernel overcomes the inherent directional specificity of conventional convolutional networks and achieves a texture recognition accuracy of 98%, thereby illustrating a viable pathway towards physics-native computing through exploiting exotic physics in quantum materials.

## Introduction

It has been a central topic to develop low-power electronic devices to address the challenges associated with computing and memory in the post-Moore era. Most existing implementations of low-power electronic devices rely on the classical charge degree of freedom of electrons as the information carrier [1]. Identifying the quantum characteristics of electrons as a new information carrier can provide a promising pathway to realize low-power electronics with functionality and performance far beyond traditional electronics [2-7]. Quantum geometry [8-10]—the intrinsic geometric structure of Bloch wavefunctions in momentum space—offers a compelling physical platform to address this challenge. Fundamental tensors of quantum geometry, such as the Berry curvature and quantum metric, not only dictate topological phases and nonlinear transport responses [11-18] but also possess distinctive attributes as "information primitives" [19-22]. Unlike charge, they encode information in the electronic phase, inherently supporting non-volatility; their control can be realized through polarization switching rather than charge transport, promising ultra-low energy operation. Higher-order quantum geometric tensors, *i.e.*, Berry curvature dipole (BCD) or Berry connection polarizability (BCP), can coexist as independent information channels, providing a physical basis for multi-level and parallel information processing.

Although remarkable progress has been made in exploring quantum geometric phenomena—particularly through the discovery of NLAHE—research has remained largely confined to proof-of-principle observations and the understanding of underlying mechanisms [23-35]. A functional demonstration of information-processing devices based on programmable quantum geometric states is still lacking. The critical step forward is to realize non-volatile electrical control of these states. If a single programming operation, such as ferroelectric polarization switching, could generate distinct responses in different harmonic orders of NLAHE within the same device, it would establish a "one-to-many" mapping between control signals and information outputs. Such a paradigm would overcome the intrinsic energy-efficiency limits of charge-based approaches and open the way toward quantum in-memory computing. However, it has been challenging to achieve nonvolatile control of quantum geometric states, because the quantum geometry and its typical transport manifestation, NLAHE, are mostly reported in metallic materials [23-35], in which ferroelectric fields fundamentally can't coexist [36-38].

The coexistence of ferroelectricity, metallicity, and significant quantum geometry in few-layer $WTe_2$ [22-24,39-42] makes it a unique platform to explore quantum-geometry-based information processing. Previous studies have demonstrated both theoretically and experimentally that $WTe_2$ holds layer-parity-dependent BCD that switches sign in odd-layer $WTe_2$ upon ferroelectric switching, thus inducing a sign switch on the second-order NLAHE at low temperature [19,43]. However, the correlation between second- and third-order NLAHE upon ferroelectric switching, especially at room temperature, has not been studied, which is a key pillar of quantum-geometry-enabled information processing.

Here we demonstrate, for the first time, non-volatile, deterministic, and correlated electrical switching of the second- and third-order NLAHE in few-layer $WTe_2$ through ferroelectric polarization switching. The switching is robust at room temperature with remarkable cycling and retention stability. Moreover, by exploiting the opposite switching polarities of the two high-harmonic signals, we construct a complementary convolution kernel based on a single NLAHE device. This kernel resolves the direction-specific constraints of conventional convolutional networks and enables a texture recognition accuracy of 98%. Our results move beyond the mere observation of quantum geometric phenomena to establish their functional role as a programmable information-processing core, paving the way for quantum in-memory computing technologies.

## Results and Discussion

### Ferroelectric NLAHE in trilayer $WTe_2$

To systematically study the NLAHE in $WTe_2$, we fabricate both odd- and even-layer $WTe_2$ devices with trilayer, tetralayer, and pentalayer $WTe_2$, namely devices D1-D4 (D1, D4: trilayer, D2: pentalayer, D3: tetralayer). In the main text, we show the data from device D1 at low temperature and device D2 at room temperature, and additional data from other devices are available in Supplementary Note 3-5. To probe the high-order NLAHE in $WTe_2$, we use the standard four-probe technique by applying a longitudinal a.c. current $I_{xx}^{\omega}$ and probe the first-order longitudinal voltage ($V_{xx}^{\omega}$), second- ($V_{xy}^{2\omega}$) and third-order ($V_{xy}^{3\omega}$) Hall voltages simultaneously. As shown in Figure 1a, when an a.c. current $I^{\omega}$ passes through few-layer $WTe_2$, the intrinsic BCD gives rise to the second-order NLAHE ($V_{xy}^{2\omega}$). The injected current also induces a field-induced BCD through the BCP tensor and generates the third-order NLAHE ($V_{xy}^{3\omega}$). When interlayer sliding induced ferroelectric transitions in odd-layer $WTe_2$, for example, trilayer (left panel of Figure 1a), the intrinsic BCD switches sign while the field-induced BCD only changes its magnitude (right panel of Figure 1a). As a result, it's possible to use ferroelectric switching to control the $V_{xy}^{2\omega}$ and $V_{xy}^{3\omega}$ in $WTe_2$ simultaneously, but in different ways.

To independently control the ferroelectricity and doping density in $WTe_2$, and their effects on NLAHE, we fabricate dual-gate Hall-bar devices with few-layer $WTe_2$ sandwiched between top and bottom graphite / boron nitride (hBN) gates, with top and bottom hBN thicknesses being $d_t$ and $d_b$, respectively (See Supplementary Table 1 for detailed AFM characterization results). Gate voltages $V_t$ and $V_b$ can be independently applied through the graphite gate on the $WTe_2$ flake, the gate-induced vertical electrical field in $WTe_2$ is given by $E_{\perp} = \frac{(-V_t/d_t + V_b/d_b)}{2}$ and the net electron doping density is $n_e = \frac{\varepsilon_{hBN}\varepsilon_0(V_t/d_t + V_b/d_b)}{e}$, the relative permittivity of hBN is adopted as $\varepsilon_{hBN} \approx 3.5$. By applying dual gate voltages according to the relationships given by the equations, we can independently control over electrical field and doping in $WTe_2$.

The crystallography dependence of $V_{xy}^{2\omega}$ in $WTe_2$ shows the signal maximizes when current is applied along the symmetry broken crystal a-axis and disappears when current is applied along the b-axis [24]. The $V_{xy}^{3\omega}$, however, is forbidden when current is strictly along the a- or b-axis but reaches maximum when the current is off from a-axis by ~15° [12,25]. To ensure the coexistence of $V_{xy}^{2\omega}$ and $V_{xy}^{3\omega}$ in $WTe_2$, we fabricated our device with the longitudinal direction, along which current is injected, slightly off the crystal a-axis. Based on our previous work and STEM measurement results [44], we first select our $WTe_2$ flakes with long, straight edge, where the a-axis tend to align. During fabrication, the long, straight edge of $WTe_2$ is slightly rotated to create a small angle between the crystal a-axis and current direction. As a result, we simultaneously observe the $V_{xy}^{2\omega}$ and $V_{xy}^{3\omega}$, which scale to quadratically and cubically to the applied current, respectively (Figure 1b). It is worth noting that the $V_{xy}^{3\omega}$ is nearly one order of magnitude larger than $V_{xy}^{2\omega}$, which agrees well with the crystal orientation dependence of $V_{xy}^{2\omega}$ and $V_{xy}^{3\omega}$. Further verification measurements are performed to exclude any extrinsic contributions besides NLAHE (Supplementary Note 2).

$E_{\perp}$ is first swept to maximum along $+z$ / $-z$ direction and returned to zero, thus presetting the ferroelectric polarization to P↑ / P↓ state. The ferroelectricity of $WTe_2$ is then confirmed by measuring the vertical electrical field dependence of the first-order longitudinal resistance $R_{xx}^{\omega}$. Figure 1c shows the butterfly-shaped hysteresis loop in $R_{xx}^{\omega}$ under sweeping $E_{\perp}$, which resembles previous work [39] and confirms the ferroelectricity in our sample. We note that the ferroelectric transition in our sample consists of multiple jumps, which is due to the formation of multi-domains in $WTe_2$ and is commonly seen in 2D ferroelectric systems, including $WTe_2$ [39,41]. We also note that the resistance jumps representing the ferroelectric transition are asymmetric with respect to $E_{\perp} = 0$, which can be ascribed to a small built-in electrical field during device fabrication [41] or charge trapping effects at interfaces [39,45].

Next, we investigate the electrical field dependence of $V_{xy}^{2\omega}$ and $V_{xy}^{3\omega}$. Both the $V_{xy}^{2\omega}$ (Figure 1d) and $V_{xy}^{3\omega}$ (Figure 1e) show hysteresis behavior and sudden signal jumps corresponding to ferroelectric transitions, clearly indicating a direct coupling of ferroelectricity to both the second- and third-order NLAHE. Both $V_{xy}^{2\omega}$ and $V_{xy}^{3\omega}$ show a monotonic dependence on electrical field, with $V_{xy}^{2\omega}$ showing a positive slope and $V_{xy}^{3\omega}$ showing a negative slope. It is worth noting that the sign of $V_{xy}^{2\omega}$ is opposite at both ends of electrical field, corresponding well to the theory of ferroelectric NLAHE in odd-layer $WTe_2$, where the BCD switches sign for the two ferroelectric polarizations [19,43]. However, when the field is at 0, $V_{xy}^{2\omega}$ remains the same sign, which can be ascribed to the contribution of the scattering [15] mechanism prominent in few-layer $WTe_2$, such as skew scattering [24]. The scattering mechanism causes an additional offset on the $V_{xy}^{2\omega}$ and thus a negative shift of the signal. The sign reversal behavior is also observed at low temperature on all devices with odd-layer $WTe_2$ (Supplementary Figure 8) and is not observed in the tetralayer $WTe_2$ sample, where the layer number is even (Supplementary Figure 9). On the contrary, even though $V_{xy}^{3\omega}$ also shows a large dependence on field, its sign does not switch throughout the sweep, which is also confirmed in all our samples (Supplementary Figure 8, 9). $V_{xy}^{3\omega}$ originates from the field-induced BCD, meaning its sign is fixed by the applied in-plane a.c. electrical field and thus cannot be reversed by ferroelectric switching. However, there is longitudinal conductance change associated with ferroelectric switching and changes nonlinear Hall conductivity. As a result, ferroelectric switching only leads to magnitude change in $V_{xy}^{3\omega}$.

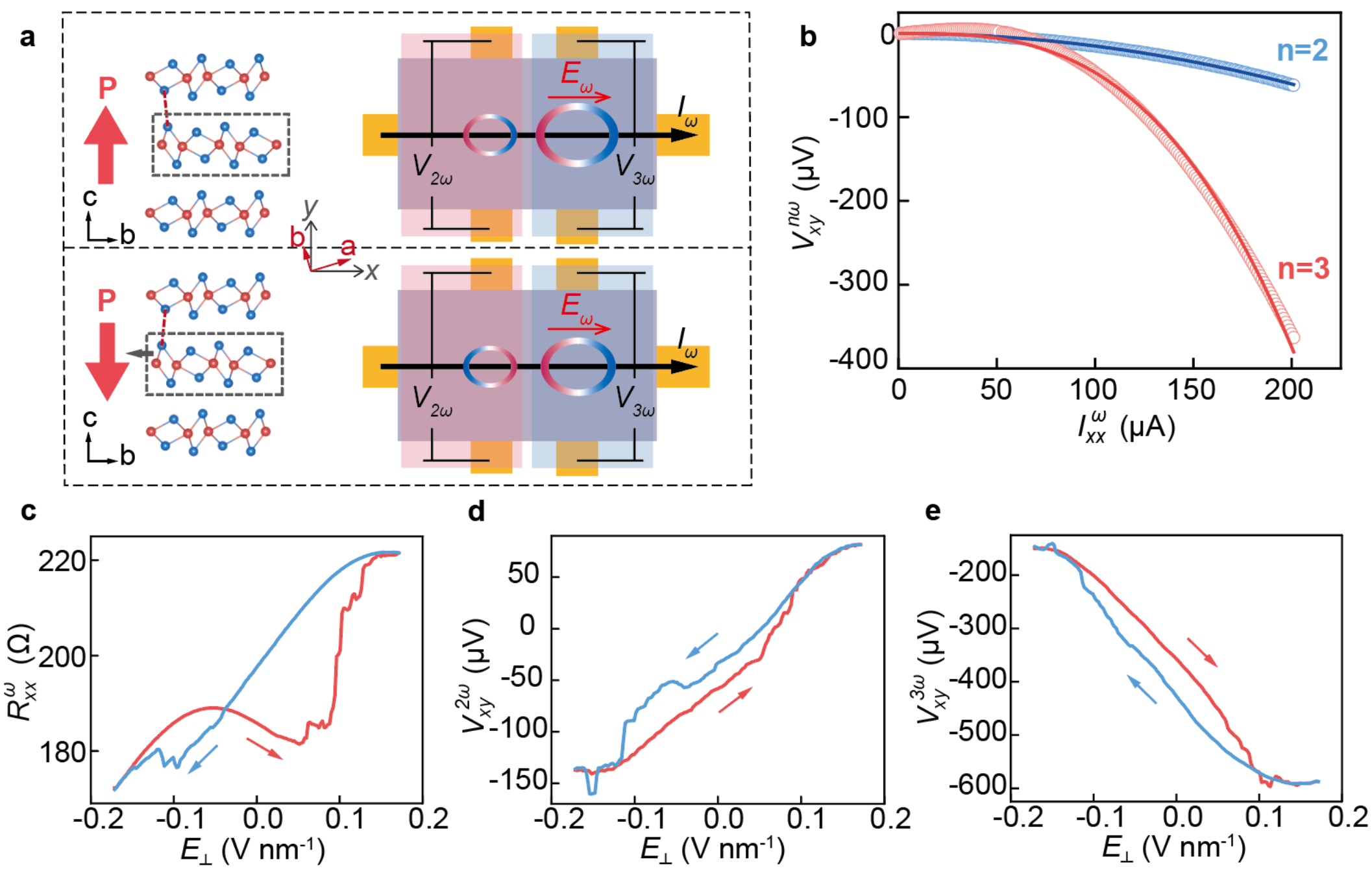

**Figure 1 | Ferroelectricity and ferroelectric nonlinear Hall effect (NLAHE) in trilayer $WTe_2$. a,** Left: Interlayer translation of trilayer $WTe_2$ during ferroelectric switching. Right: ferroelectricity modulated second- and third-order nonlinear Hall voltages. The intrinsic Berry curvature and the field-induced Berry curvature, shown as colored ellipses in the right panel, show a dipolar distribution in momentum space. During ferroelectric switching, only the intrinsic Berry curvature dipole (left) switches sign, while the field-induced Berry curvature dipole (right) only changes magnitude. P denotes ferroelectric polarization, $I_\omega$ denotes applied a.c. current and $E_\omega$ denotes a.c. electrical field across the sample. $V_{2\omega}$ and $V_{3\omega}$ denote the second- and third-order nonlinear Hall responses, respectively. **b,** Nonlinear Hall responses $V_{xy}^{n\omega}$ as a function of applied current $I_{xx}^{\omega}$ of trilayer $WTe_2$ at 2 K. Experiment data are plotted as blue ($V_{xy}^{2\omega}$) and red ($V_{xy}^{3\omega}$) dots and solid lines are the quadratic and cubic fits to the data, respectively. **c-e,** Vertical electrical field dependence of first-order longitudinal

resistance $R_{xx}^{\omega}$, $V_{xy}^{2\omega}$, $V_{xy}^{3\omega}$ at 2 K. Signals of opposite sweeping directions are separated by blue and red colors, and their sweeping directions are indicated by colored arrows.

**Doping modulation and scaling analysis of NLAHE**

Next, we perform the dual-gate mapping of $R_{xx}^{\omega}$, $V_{xy}^{2\omega}$ and $V_{xy}^{3\omega}$ to comprehensively showcase the dependence of the NLAHE signals on the electrical field and doping. In the dual gate mapping, the top gate is swept in two opposite directions, denoted as positive (→) and negative (←). The bottom gate is swept with a certain relationship as $V_b = -\frac{d_b}{d_t}V_t + V_{offset}$. The coefficient between the top and bottom gate voltage $-\frac{d_b}{d_t}$ is used to keep the net doping stable throughout the sweep ($\Delta n_e = 0$) and the offset of the bottom gate $V_{offset}$ is applied to add an external doping ($n_\mathrm{e} = \frac{\varepsilon_{hBN}\varepsilon_0(V_{offset}/d_b)}{e}$) on the sample.

Figure 2a plots the differences of $R_{xx}^{\omega}$ in the two directions as $\Delta R_{xx}^{\omega} = R_{xx,\rightarrow}^{\omega} - R_{xx,\leftarrow}^{\omega}$. A clear separation of red and blue regions is observed, confirming the butterfly-shaped loop observed in field sweep measurements. The dual gate mappings of $R_{xx}^{\omega}$, $V_{xy}^{2\omega}$ and $V_{xy}^{3\omega}$ with single sweeping direction in device D1 are available in Supplementary Figure 5.

We measure the doping dependence of $V_{xy}^{2\omega}$ and $V_{xy}^{3\omega}$ by sweeping dual-gate voltages along the black dashed line ($E_\perp = 0$) in the $\Delta R_{xx}^{\omega}$ mapping. Both $V_{xy}^{2\omega}$ and $V_{xy}^{3\omega}$ show a large dependence on external doping (Figure 2b, 2c). For the doping dependence of $V_{xy}^{2\omega}$, the difference of signal between the two ferroelectric states is more obvious under either large electron or hole doping, and a large separation of $V_{xy}^{2\omega}$ is observed at the hole doping end. The doping dependence of $V_{xy}^{2\omega}$ is consistent with the symmetry-dependent feature of BCD in $WTe_2$, where the non-centrosymmetric ferroelectric $T_d$ phase is preferred under large hole doping [19]. The symmetry broken crystal lattice then gives rise to a sizeable ferroelectric-coupled BCD, which in turn lead to the large separation in $V_{xy}^{2\omega}$. We note that even though both the field and doping can effectively modulate $V_{xy}^{2\omega}$, the sign switch is only observed in the field sweep measurement, thus confirming the sign switch of $V_{xy}^{2\omega}$ due to the ferroelectric-induced sign switch of BCD, and rules out the possibility of doping-induced sign switch. As for the doping dependence of $V_{xy}^{3\omega}$, the difference of $V_{xy}^{3\omega}$ between two ferroelectric states is more obvious when there is small or no external doping, and the separation of signals disappears at either high electron or hole doping end. It is also worth noting that the slope of $V_{xy}^{3\omega}$ under sweeping electrical field and doping is opposite, which rules out the possibility of uncompensated doping modulation of $V_{xy}^{3\omega}$ in the field sweep measurements. Even though both $V_{xy}^{2\omega}$ and $V_{xy}^{3\omega}$ show good modulation by external doping, the difference in their doping and electrical field dependence indicates their distinct origins. This will further enable the possibility to distinguish multiple nonlinear readouts in our quantum geometry memory, not only by the harmonic order, but also by the distinctive dependence features of different readouts.

We further validate the origin of the observed second- and third-order NLAHE by performing scaling analysis based on temperature-dependent data. Figure 2d shows the temperature dependence of longitudinal conductance σ, which in general shows a metallic feature with higher conductance at low temperature, corresponding to the semi-metallic nature of trilayer $WTe_2$. In few-layer $WTe_2$, the nonlinear conductivity can be fitted by $E_{xy}^{n\omega}/(E_{xx}^{\omega})^n = C_1\sigma^2 + C_0 \approx \frac{\chi}{r\sigma}$, where $\chi$ is nonlinear Hall susceptibility and $r \approx 0.3$ is the resistance anisotropy of $WTe_2$ [24]. Since $\sigma \propto \tau$, the $C_1$ coefficient represents the mechanism with $\tau^3$ dependence, such as the skew scattering mechanism; whereas the $C_0$ coefficient represents the intrinsic BCD or BCP mechanism that is proportional to $\tau$ [15,24]. Figure 2e shows the scaling relationship of the second-order NLAHE in trilayer $WTe_2$. We observe a non-monotonic dependence on conductance, signaling a mixture of different mechanisms. At low temperatures (high conductance end), the second-order nonlinear conductance has a small slope with a large intercept, meaning that the intrinsic BCD contribution dominates. The large separation of signals between the two ferroelectric polarizations also confirms that the intrinsic symmetry-dependent BCD is the main contribution to the second-order NLAHE. However, at high temperatures (low conductance end), the second-order nonlinear conductance shows a linear relationship with $\sigma^2$ with a much larger slope, meaning that the effect of scattering becomes obvious above 100 K. However, the non-zero intercept for both polarizations at the low conductance end shows that the intrinsic BCD contribution, as well as the ferroelectric modulation, exists even at room temperature. Figure 2f shows the scaling analysis of third-order nonlinear conductance, where a fine linear relationship is observed at low temperatures (high conductance end), proving a BCP-dominated mechanism [25]. The deviation of the signal at high temperature can be ascribed to the temperature-driven Fermi level shift [46,47] that becomes more pronounced at high temperatures. The scaling analysis of device D3 with even-layer $WTe_2$ shows similar behavior, as discussed in Supplementary Figure 11.

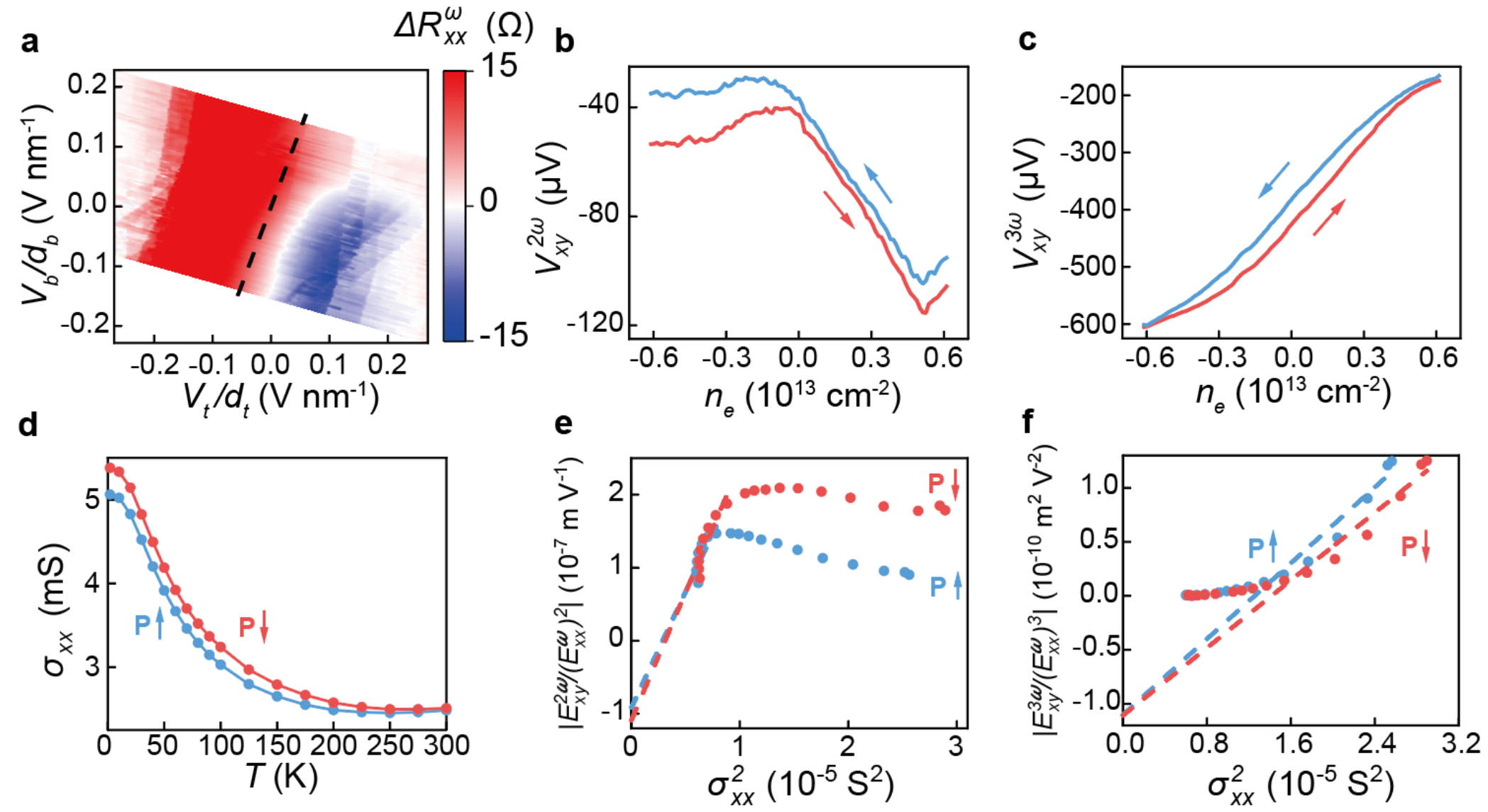

**Figure 2 | Doping modulation and scaling relationship of nonlinear responses**. **a,** Dual gate mapping of first-order longitudinal resistance difference $\Delta R^{\omega}_{xx}$ between positive (→) to negative (←) sweeps at 2 K. The sign of $\Delta R^{\omega}_{xx}$ is represented by red (positive) and blue (negative) colors, respectively. The black dashed line denotes $E_{\perp} = 0$ trace. $V_b$ ($V_t$) and $d_b$ ($d_t$) denote the gate voltage and hBN thickness of top (bottom) gate, respectively. The doping dependence of NLHAE voltages **b,** $V^{2\omega}_{xy}$ and **c,** $V^{3\omega}_{xy}$ of two ferroelectric polarizations along the $E_{\perp} = 0$ trace at 2 K. **d,** The evolution of conductivity σ of two ferroelectric polarizations at various temperatures from 2 K to 300 K. The nonlinear Hall conductivity $E^{n\omega}_{xy}/(E^{\omega}_{xx})^n$ as a function of $\sigma^2_{xx}$ for **e,** second-order ($n = 2$) and **f,** third-order responses ($n = 3$), respectively. The dashed lines are the corresponding linear fits to the data.

## Room-temperature stability tests of pentalayer $WTe_2$

While the ferroelectric switching in $WTe_2$ has been shown with remarkable stability at low temperature (100 K) [41], the performance at room temperature has not been tested. Moreover, stable ferroelectric switching does not necessarily lead to the stable switching of NLAHE, due to the extra mechanisms involved. Hence, it's important to investigate the switching stability of NLAHE at room temperature before discussing its possible applications. First, we observe ferroelectric switching behavior of all signals persists at room temperature (Figure 3a-c for device D2 and Supplementary Figure 6 for device D1). The sign reversal behavior of $V_{xy}^{2\omega}$ is not observed at room temperature (Figure 3b), which could also be due to the shift of the Fermi level at higher temperatures and the offset from the extrinsic scattering contribution. We note that the dependence of $V_{xy}^{2\omega}$ and $V_{xy}^{3\omega}$ on sweeping field shows opposite trend in the small field range, *i.e.* the slope of the positive to negative sweep (blue line) in $V_{xy}^{2\omega}$ is positive at field close to zero, but for $V_{xy}^{3\omega}$ the slope is negative. Moreover, this sign reversal feature is attributed to the electrical field induced changes in the intrinsic BCD and BCP contribution and is confirmed by our calculation results (Supplementary Note 8).

To investigate the device stability of nonvolatile switching of first-, second-, and third-order responses in $WTe_2$, we adopt the following protocol to perform cycling and retention measurements. Two voltage pulses are applied simultaneously through both top and bottom gates to provide a pulse electrical field of 0.39 V nm$^{-1}$ along $+z$ $(-z)$ direction to set the polarization to the P↑ (P↓) state, then the electrical field is withdrawn, and first-order longitudinal, second-, and third-order Hall responses are collected simultaneously. For the cycling measurements, $WTe_2$ is alternatively set to P↑ and P↓, and signals of all three harmonic orders are collected after each polarization setting, when the vertical electrical field is returned to zero. For the retention measurements, $WTe_2$ is first set to either P↑ or P↓ state, and the vertical electrical field is returned to zero. Then the source meters used to provide gate voltages are removed from the circuit and a.c. response of all three harmonic orders are collected over a period of time.

Figure 3d-f shows the cycling stability of $R_{xx}^{\omega}$, $V_{xy}^{2\omega}$, $V_{xy}^{3\omega}$. In a total of $1.4 \times 10^4$ cycles, all three signals show good stability after the switching measurement. This is further validated by the retention measurement, which is conducted right after the cycling measurement. For the retention measurement, all three signals at both ferroelectric polarizations remain stable for $10^5$ seconds after switching (Figure 3g-i). The cycling and retention measurement results clearly indicate that the ferroelectric-coupled NLAHE in few-layer $WTe_2$ can be electrically and non-volatilely manipulated with great fatigue resistance and stability, paving the way for its application as a novel memory device with quantum geometric protected multiple readouts. The stability measurement results of device D1 are shown in Supplementary Figure 7.

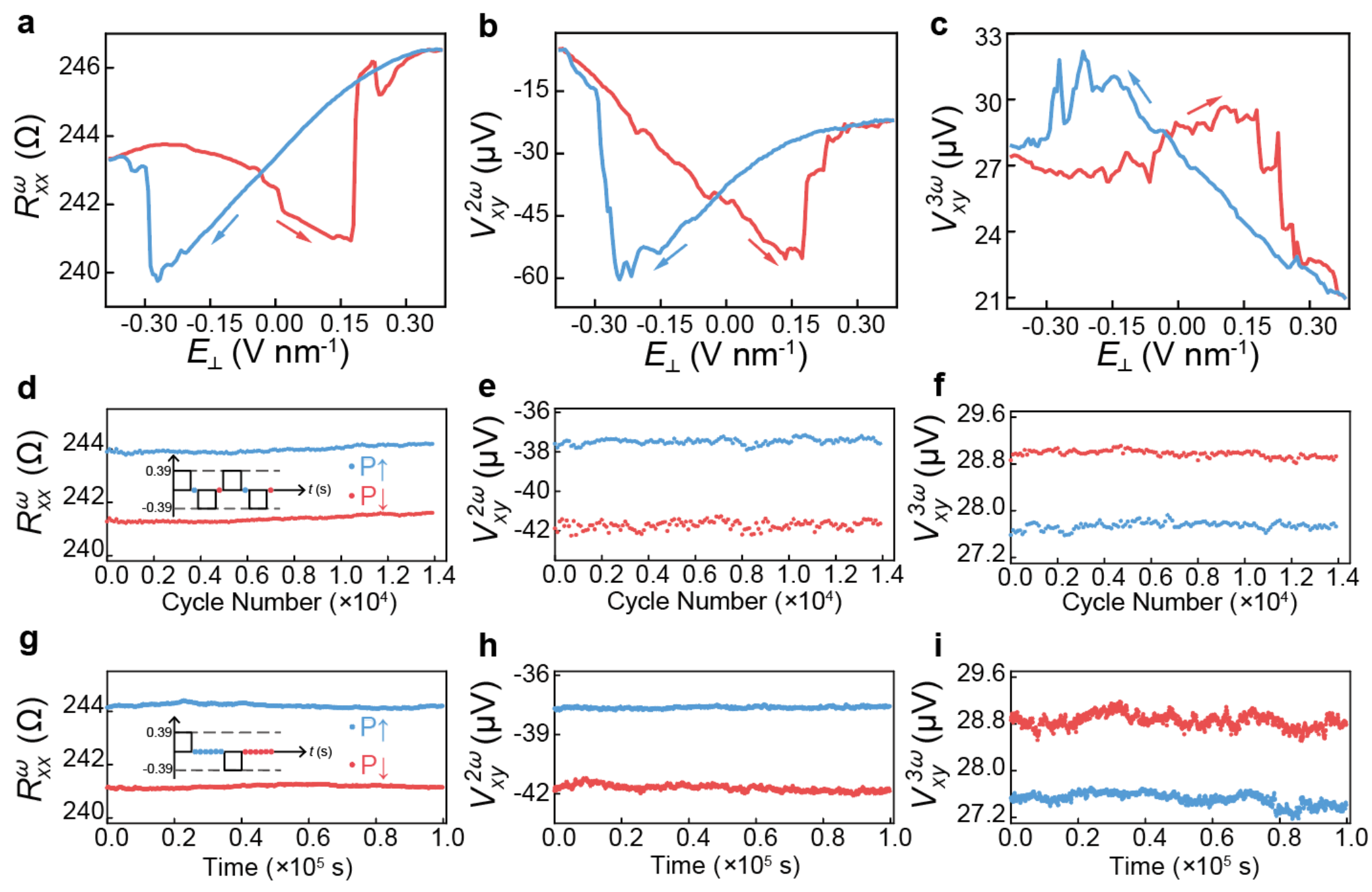

**Figure 3 | Stability tests in pentalayer $WTe_2$ device D2 at room temperature. a-c,** Electrical field dependence of $R_{xx}^{\omega}$, $V_{xy}^{2\omega}$, $V_{xy}^{3\omega}$ of device D2 at room temperature. **d-f,** Switching stability results of $R_{xx}^{\omega}$, $V_{xy}^{2\omega}$, $V_{xy}^{3\omega}$ of device D2 with $1.4 \times 10^4$ consecutive field sweep cycles using 0.39 / -0.39 V nm$^{-1}$ to set the polarization to P↑ / P↓. Inset of **d** shows the schematic illustration of the switching measurement. **g-i,** Retention measurement results of $R_{xx}^{\omega}$, $V_{xy}^{2\omega}$, $V_{xy}^{3\omega}$ of device D2 in $10^5$ seconds after setting the polarization to P↑ / P↓ with 0.39 / -0.39 V nm$^{-1}$ field. Inset of **g** shows the schematic illustration of the retention measurement.

## Complementary convolution kernel based on NLAHE in $WTe_2$

By exploiting the simultaneous and opposite responses of $V_{xy}^{2\omega}$ and $V_{xy}^{3\omega}$ in $WTe_2$ to ferroelectric polarization, we realize a complementary convolutional kernel that enables orientation-complementary feature extraction of texture pattern at the device level. We notice that the room-temperature $V_{xy}^{2\omega}$ scales linearly under small electrical field change $\Delta E_{\perp}$ for both polarization states, enabling a multiplicative operation $V_{\mathrm{xy}}^{2\omega} = w \cdot \Delta E_{\perp}$, where the weight $w$ is set by the ferroelectric polarization state (positive $w$ for P↑ state and negative $w$ for P↓ state). In contrast, $V_{xy}^{3\omega}$ yields the opposite computational effect under identical polarization and configuration (negative $w$ for P↑ state and positive $w$ for P↓ state), performing a multiplication $V_{\mathrm{xy}}^{3\omega} = -w \cdot \Delta E_{\perp}$ with opposite sign. This unique property allows a single programmed weight distribution to generate complementary computational outputs through different harmonic readouts.

Based on this mechanism, we construct a 3 × 3 ferroelectric NLAHE device array as a convolutional kernel (Figure 4a). Each device has independently addressable gates for programming polarization states and applying encoded input voltages. The electrodes along the $xy$ directions are connected to a common source line clamped at 0 V for readout. During programming, each device is set to up or down polarization, corresponding to binary kernel weights (−1 and +1). During convolution, input voltages are applied to the pixel devices, which perform multiply-accumulate operations under the ferroelectric polarization response of the NLAHE, and readouts at different harmonic orders produce the convolution outputs. Unlike conventional resistive-memory–based in-memory convolution kernels, which generate a single output per weight configuration, the proposed complementary kernel produces paired outputs of opposite polarity under identical weights. Specifically, the second-order readout yields one set of convolution results, while the third-order readout produces the opposite results.

For typical edge-detection kernels, the second- and third-order readouts correspond to feature extraction along a given direction and its opposite, respectively. By combining these outputs, diagonal-orientation-complete (four-quadrant) edge features can be extracted without introducing additional kernels. We further configure the 3 × 3 array into two different directional kernels (Figure 4b). Kernel 1 is used for extracting edges from top-left to bottom-right, while Kernel 2 is configured to extract edges from bottom-left to top-right. Using a "cracker" image as an example, the second-harmonic readouts reveal edges aligned with the programmed kernel directions (top panel of Figure 4c), whereas the third-order readout produces edge responses along the opposite diagonal directions (bottom panel of Figure 4c). Subtracting the second- and third-order outputs produces enhanced edge maps with improved directional coverage.

To evaluate the functional impact of complementary edge extraction, we implement a convolutional neural network comprising a single convolution layer, followed by pooling and a fully connected layer (Figure 4d), and apply it to the KTH-TIPS texture recognition dataset [48,49]. (See Methods for details of network simulation). The KTH-TIPS dataset contains ten texture classes under varying illumination, orientation, and scale, including natural materials (sandpaper, crumpled aluminum foil, styrofoam, sponge), fabrics (corduroy, linen, cotton), and structures such as brown bread, orange peel, and cracker (left panel of Figure 4d). Under identical network architecture and training conditions, the complementary convolution achieves higher recognition accuracy than conventional in-memory convolution, in which each weight configuration yields only a single directional response unless additional kernels are introduced. This performance gain arises from more complete edge feature extraction rather than increased model complexity. As a result, edge responses in the complementary orientations (bottom-right to top-left and top-right to bottom-left) remain absent even when two directional kernels are used, leading to incomplete edge representations and a reduced recognition accuracy from 98% to 89% (Figure 4e and 4f). To evaluate the impact of device non-idealities, experimentally measured input-output transfer characteristics and noise levels were incorporated into the simulation (Supplementary Figure 12). Under these conditions, the recognition accuracy of the complementary convolution slightly decreases to 94%, while remaining significantly higher than that of the conventional scheme. Importantly, achieving comparable orientation coverage with linear unidirectional kernels would require additional kernels, leading to increased hardware overhead, latency, and programming complexity.

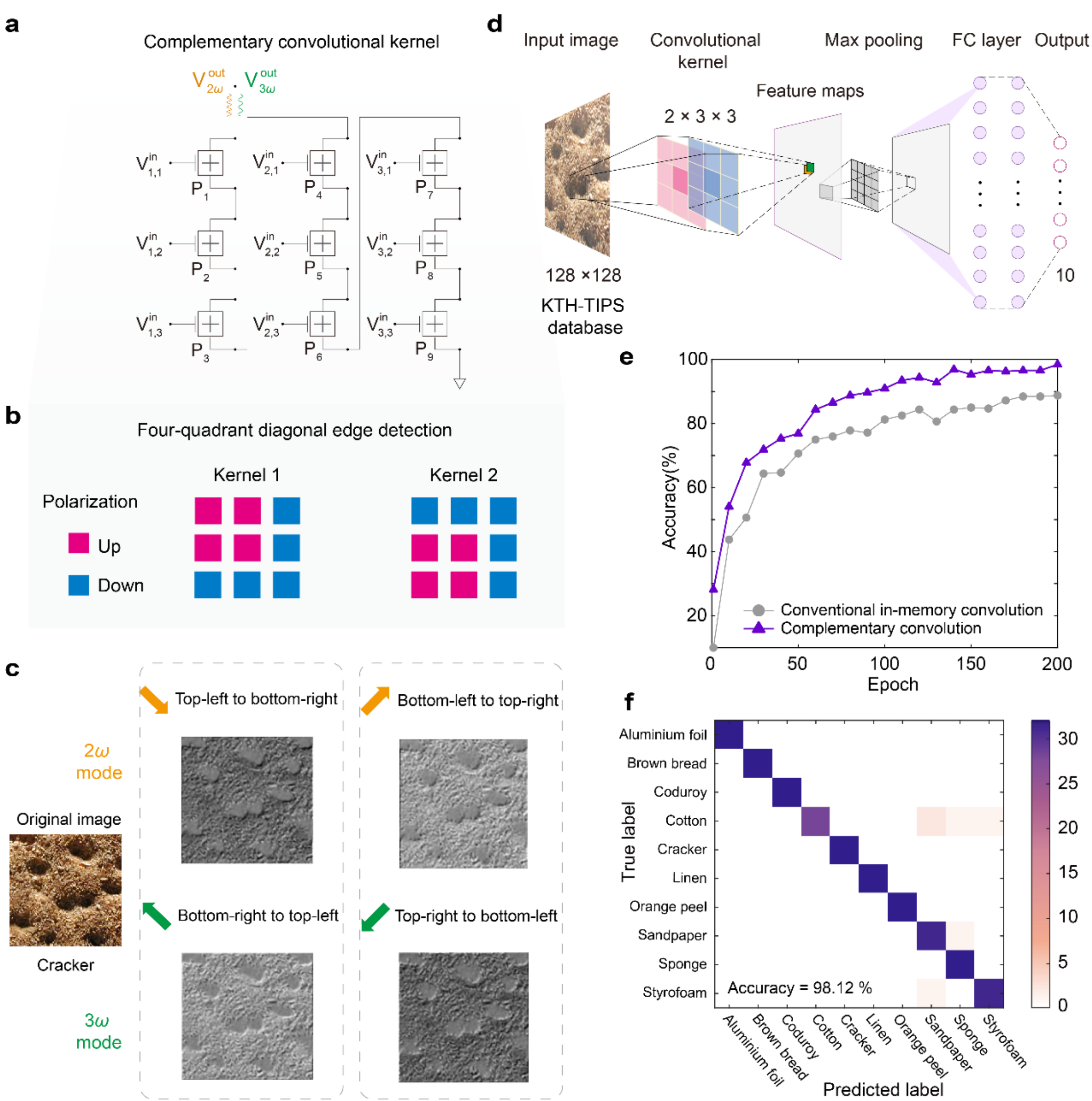

**Figure 4 | Complementary convolution for edge detection and texture recognition.**

**a**, Schematic of the complementary convolution kernel, which performs convolutions in opposite directions by switching between the second- and third-harmonic readout frequencies. Here, $P_n$ $(n = 1 - 9)$ denotes independent ferroelectric polarization states of the nine devices in the 3 × 3 array and $V_{n,m}^{in}$ $(n, m = 1,2,3)$ denotes small input voltage on each device at $(n, m)$. $V_{2\omega}^{out}$ and $V_{3\omega}^{out}$ denote the second- and third-harmonic output voltages, respectively. **b**, The configured four-quadrant diagonal edge detection kernels: Kernel 1 detects edges along top-left to bottom-right, and Kernel 2 along bottom-left to top-right. **c**, Computational results of the complementary convolution kernels: Left, output of Kernel 1; Right, output of Kernel 2. An example input image of a “cracker” from the KTH-TIPS dataset is shown (used under the CC BY-NC 4.0 license). **d**, Convolutional neural network architecture used for KTH-TIPS image recognition, including convolutional, pooling, and fully connected (FC) layers. Representative texture image is from the KTH-TIPS dataset (used under the CC BY-NC 4.0 license). **e,** Simulated recognition accuracy on the KTH-TIPS database over 200 epochs, comparing complementary and conventional convolutional processing under ideal conditions and with device non-idealities. **f,** Confusion matrix for the test dataset under ideal simulation conditions.

These results demonstrate that the proposed complementary convolution operates as a hardware-efficient convolutional primitive based on high-order NLAHE devices, which feature switchable quantum geometric states and inherently generate both forward and reverse edge responses through harmonic readouts. This capability allows orientation-complementary feature extraction without duplicating weights or introducing additional kernels, thereby enabling weight sharing across orientations, reducing multi-kernel redundancy, and eliminating extra memory requirements. The resulting complementary features further enhance robustness to mirrored and symmetric patterns, while maintaining high computational efficiency.

To conclude, we have successfully demonstrated that quantum geometry in few-layer $WTe_2$, manifested as nonlinear anomalous Hall effects, can be electrically switched at room temperature. The switching is nonvolatile and robust, with quantum geometric states retaining $10^5$ seconds and switching $1.4 \times 10^4$ cycles without degradation. Finally, we propose a first-of-its-kind complementary convolution network based on the switchable quantum geometric states. The complementary responses of different quantum geometric states enable in-memory computing to achieve texture recognition with 98% accuracy, well exceeding conventional convolution implemented with traditional in-memory computing devices. Our demonstrations highlight the potential of quantum geometry as a new degree of freedom for information encoding and processing.

## Methods

**Device fabrication and characterizations.** All crystals in this study were purchased from HQ Graphene and used as received. Graphite and hBN flakes were prepared by mechanical exfoliation onto $SiO_2$ / Si substrate with $SiO_2$ thickness of 285 nm. The flakes were examined by an optical microscope to select the flakes with a clean surface and appropriate thickness in ambient conditions. hBN / graphite stack was prepared by polycarbonate (PC) / polydimethylsiloxane (PDMS) stamp-assisted dry-transfer technique inside a $N_2$ glovebox with $O_2 < 0.2$ ppm and $H_2O < 0.1$ ppm. The stack was examined carefully under a microscope in ambient conditions to confirm a clean graphite / hBN interface. After spin-coating polymethyl methacrylate (PMMA), standard electron beam lithography (EBL) was performed on the hBN / graphite stack, followed by electron beam deposition of 5 nm Ti / 25 nm Au to finish the electrode fabrication.

$WTe_2$ flakes were mechanically exfoliated on $SiO_2$ / Si substrate with the same $SiO_2$

thickness inside the same $N_2$ glovebox mentioned above. Suitable $WTe_2$ flakes were examined by microscope inside the $N_2$ glovebox and selected based on their optical intensity contrast profiles. The crystal a-axis of $WTe_2$ flakes was determined as the long, straight edges of the flakes. Finally, a graphite / hBN / $WTe_2$ stack was assembled and transferred onto the pre-patterned electrodes using the same dry-transfer technique as mentioned above inside the same $N_2$ glovebox.

The thickness of both top and bottom hBN layers was characterized by atomic force microscopy (AFM). A summary of thickness information and characterization results of all devices can be found in Supplementary Note 1.

**Electrical measurement.** Electrical measurements were performed in an Oxford Teslatron cryostat down to 2 K. Keithley 6430 and Keithley 2401 source meters were used to provide D.C. gate voltages to the top and bottom gates, respectively. An SR830 lock-in amplifier was used to provide a.c. current through $WTe_2$ along the longitudinal direction, while the first-order longitudinal voltage, second- and third-order Hall voltages were collected simultaneously by three inter-connected SR830 lock-in amplifiers. The frequency of the a.c. current applied was 13.317 Hz, unless otherwise stated. The phases for first- and third-order responses were 0° and ±180°, while the phase of the second-order response was ±90°, consistent with our expectations.

**Implementation of KTH-TIPS recognition.** To evaluate the convolution device, we simulated a network for KTH-TIPS texture recognition, consisting of a single convolutional layer, a pooling layer, and fully connected layers. The dataset includes ten texture classes: real materials (sandpaper, crumpled aluminum foil, styrofoam, sponge), fabrics (corduroy, linen, and cotton), and natural structures (brown bread, orange peel, and cracker). Each class contains 81 images across 9 scales, of which 49 images were used for training and the remaining for testing. All images were resized to 128 × 128 pixels prior to training. To account for device non-idealities, Gaussian noise corresponding to the residual-based noise levels extracted from the transfer-curve fitting of the second- and third-harmonic responses was added during the convolution process. During forward propagation, each input image was processed by two bidirectional convolution kernels. The convolution results from the second- and third-order responses were subtracted to generate a 2 × 128 × 128 feature map. This feature map was then processed with a 1 × 1 channel expansion convolution followed by 8 × 8 max pooling, yielding a 4 × 16 × 16 feature map. The pooled feature map was reshaped and fed into a fully connected layer of size 1024 × 256 × 10 to produce classification outputs. Training was performed for 200 epochs, and the resulting confusion matrix for the test dataset was obtained.

**Data Availability**

Relevant data supporting the key findings of this study are available within the article and the Supplementary Information file. The data that support the plots within this paper are available at the Repository DR-NTU (DATA): https://researchdata.ntu.edu.sg/dataset.xhtml?persistentId=doi:10.21979/N9/F4VB6D

**Funding Statement**

P. S. acknowledges start-up grant support from Nanyang Technological University and Singapore Ministry of Education Academic Research Fund Tier 2 (MOE-T2EP50122-0017) and Tier 1 (RG113/21, RG130/22, RG76/25) projects. F. M. and S. J. L. would like to acknowledge support from AIQ foundation. This work was supported in part by the National Key R and D Program of China under grant 2023YFF1203600 (S. J. L.), the National Natural Science Foundation of China (62404099 (P. F. W.), the Leading-edge Technology Program of Jiangsu Natural Science Foundation (BK20232004 (F. M.)), the Natural Science Foundation of Jiangsu Province (BK20233001), the AI & AI for Science Project of Nanjing University (14380240, 14380242, and 14380005), the Fundamental Research Funds for the Central Universities (14380227, 14380247, 14380250).

**Author Contributions Statement**

P. S. conceived and supervised the project. H. C. fabricated the devices with the help of B. L. and X. C.. R. W. performed the electrical measurements and analyzed the data. R. W. performed the AFM measurements. P. W. performed the convolution network simulation with the help of J. X., X. Z. and C. P.. Y. P. and S. A. Y. performed the first-principles calculations. R. W., P. W., Y. P., S. A. Y., S. J. L., F. M., and P. S. wrote the manuscript with input from all authors. All authors discussed the results and commented on the manuscript.

**Competing Interests Statement**

The authors declare no competing interests.